# Pressure effects on the superconductivity of HfPd$_2$Al Heusler compound: Experimental and theoretical study

B. Wiendlocha[1,*], M. J. Winiarski[2], M. Muras[1], C. Zvoriste-Walters[3], J.-C. Griveau[3], S. Heathman[3], M. Gazda[2], T. Klimczuk[2,$]

[1] AGH University of Science and Technology, Faculty of Physics and Applied Computer Science, Al. Mickiewicza 30, 30-059 Krakow, Poland

[2] Faculty of Applied Physics and Mathematics, Gdansk University of Technology, Narutowicza 11/12, 80-233 Gdansk, Poland

[3] European Commission, Joint Research Centre (JRC), Institute for Transuranium Elements (ITU), Postfach 2340, 76125 Karlsruhe, Germany

* wiendlocha@fis.agh.edu.pl
$ tomasz.klimczuk@pg.gda.pl

*Abstract*

Polycrystalline HfPd$_2$Al has been synthesized using the arc-melting method and studied under ambient pressure conditions by x-ray diffraction from room temperature up to 450$^o$C. High pressure x-ray diffraction up to 23 GPa was also performed using Diacell-type membrane diamond anvil cells. The estimated linear thermal expansion coefficient was found to be $\alpha = 1.40(3) \cdot 10^{-5}$ K$^{-1}$, and the bulk modulus derived from the fit to the 3$^{rd}$ order Birch-Murnaghan EOS (BMEOS) is $B_0 = 97(2)$ GPa. Resistivity studies under applied pressure ($p \leq 7.49$ GPa) showed a linear decrease of superconducting critical temperature with increasing pressure and the slope $dT_c/dp = -0.13(1)$ K GPa$^{-1}$. The same behavior is observed for the electron-phonon coupling constant $\lambda_{ep}(p)$ that changes from 0.67 to 0.6, estimated for $p = 0.05$ GPa and 7.49 GPa, respectively. First principles electronic structure and phonon calculation results are presented and used to estimate the magnitude of electron-phonon interaction $\lambda_{ep}$ and its evolution with pressure. Theoretical results explain the experimentally observed decrease in T$_c$ due to considerable lattice stiffening.

# I INTRODUCTION

The Heusler group of alloys, discovered one hundred years ago by Friedrich Heusler [1], are well-ordered, ternary intermetallic compounds of general composition $AT_2M$, where: *A* is generally a transition metal, *T* is a transition metal from group VIIIB to IB and *M* is typically a *sp* metal or metalloid (Sb, Bi). A prototype material of this group, $MnCu_2Al$, was the first ferromagnetic alloy not to contain any ferromagnetic elements. More than hundred Heusler alloys are known to date [2], showing a variety of interesting physical properties, such as shape memory effect [3,4], magnetic ordering [5,6], half-metallic ferromagnetism [7] or heavy fermion behavior [8,9,10,11,12], that makes them an interesting group of materials for research and applications. There are also nearly 30 known Heusler superconductors with critical temperatures in the range of few Kelvins [2,13,14,15,16,17,18,19,20], most of them having rare-earth atoms in the *A* position [7]. In two of them, $YbPd_2Sn$ [16] and $ErPd_2Sn$ [18], superconductivity and magnetic ordering (antiferromagnetism) coexist. Therefore, Heusler alloys belong to a rare class of materials bridging superconducting and magnetically-ordered compounds.

In this paper we present studies on the pressure dependence of the critical temperature in $HfPd_2Al$, which were inspired by a previous finding that in cubic Heusler phase series $HfPd_2Al$, $HfPd_2In$, $ZrPd_2Al$, $ZrPd_2In$ [(Hf, Zr)$Pd_2$(In, Al)] the critical temperature increases with decreasing lattice parameter *a*. Another studied family of compounds $YPd_2Sn$, $LuPd_2Sn$, $ScPd_2Sn$ [(Y, Lu, Sc)$Pd_2Sn$], showed inverse behavior – $T_c$ increased with increasing *a* (ref. [2]). Pressure effects on superconductivity were previously studied in $REPd_2$(Sn, Pb) alloys (*RE* = Sc, Y, Tm, Yb, and Lu) and revealed linear decreases of $T_c$ with applied pressure [15]. Negative change under pressure ($dT_c/dp < 0$) was also observed in other groups of superconductors (see, e.g. Ref. [21]).

The goal of this study was to check if a further decrease in lattice parameter in group (Hf, Zr)$Pd_2$(In, Al) would lead to higher critical temperatures. The highest $T_c$ (and the lowest *a*) value in this group is reported for $HfPd_2Al$ (3.7 K) [2], and therefore this compound was chosen for studies under applied pressure.

In order to analyze and better understand the experimental results, theoretical studies were undertaken. Using the Density Functional Theory, electronic structure and phonon calculations were performed, and the magnitude of the electron-phonon coupling (EPC) was studied as a function of external pressure. The theoretical results explain qualitatively well the pressure-induced modifications of the EPC constant $\lambda_{ep}$ and critical temperature $T_c$.

**II METHODS**

Polycrystalline samples of $HfPd_2Al$ were synthesized by arc-melting stoichiometric amounts of the elements (Hf 99.5%, Pd 99.95% and Al 99.9% – all Alfa Aesar) under a zirconium-gettered ultra pure argon atmosphere. As it was pointed out in ref. [2] the post-annealing process for the $HfPd_2Al$ compound worsens superconducting properties, and therefore an as-cast $HfPd_2Al$ sample was studied.

The purity of the product was verified by powder X-ray diffraction (PXRD) using a Philips X′pert Pro MPD with $CuK_\alpha$ radiation. The high-temperature PXRD patterns were collected up to 450°C in air and a lattice parameter for $HfPd_2Al$ at different temperatures was refined by means of the Rietveld method [22] using the FullProf 5.30 program [23]. Above approximately 450 °C the sample oxidized and therefore collection of the xrd diffraction pattern above this temperature was not continued.

The high pressure study was performed by means of in-situ x-ray diffraction. Pressure was determined using the ruby scale [24] and silicone oil was used as pressure transmitting medium for all experiments. The sample was loaded into Diacell-type membrane diamond anvil cells (MDACs) with 500 μm culet size using pre-indented Re gaskets with 200 μm diameter holes. High pressure x-ray diffraction was performed using a modified Bruker D8 x-ray diffractometer with focusing mirror optics installed on a molybdenum rotating anode source (Mo $K_{\alpha 1}$), $\lambda = 0.70926$ Å, 100 x 100 μm$^2$, coupled with a Bruker SMART Apex II Charged-Coupled Device (CCD). The MDACs were rotated through a sample angle $\Delta\Omega = \pm 2$ deg. while collecting each diffraction image. The sample to CCD distance and CCD non-orthogonality correction were calibrated using powder diffraction data from a $LaB_6$ standard and the recorded diffraction images were integrated using the ESRF FIT2D software [25].

A polycrystalline sample of $HfPd_2Al$ was extracted from the batch and polished down to thickness of 20 μm. Average dimensions of the sample were therefore 750 x 100 x 20 μm$^3$. The electrical resistance of the sample was measured by a four probe dc technique with the sample and a thin foil of lead used as manometer [26] held in a pyrophyllite gasket with a solid pressure-transmitting medium of steatite. The external pressure device was a piston-cylinder system made of nonmagnetic CuBe, with the pressure generated by two 3.5 mm diameter anvils made of low-magnetic tungsten carbide and sintered diamonds. To avoid any heating effect which would modify the $T_c$ determination the applied current was relatively low (0.5 mA). Pressure was changed at room temperature and quasi hydrostatic conditions were observed during the whole experiment.

Electronic structure calculations were performed using the Korringa-Kohn-Rostorker (KKR) method [27]. Using the so-called Rigid Muffin Tin Approximation (RMTA) [28] the electronic part of the EPC constant, i.e. McMillan-Hopfield parameters η [29,30] for each $i$ atom in the unit cell were calculated [31]:

$$\eta_i = \sum_l \frac{(2l+2)\, n_l(E_F)\, n_{l+1}(E_F)}{(2l+1)(2l+3)n(E_F)} \left| \int_0^{R_{MT}} r^2 R_l(r) \frac{dV}{dr} R_{l+1} dr \right|^2 \qquad (1)$$

where $l$ is the angular momentum number, $n_l(E_F)$ are l-decomposed densities of states (DOS) at the Fermi level at atom $i$, $n(E_F)$ is the total DOS at $E_F$ per primitive cell (DOSes are given per spin), $R_l(r)$ are normalized radial wave functions and $R_{MT}$ is the muffin tin radius (for discussion and examples of application of RMTA see refs. [31, 32, 33, 34, and references therein]. The crystal potential was constructed in the framework of the local density approximation, using the von Barth and Hedin [35] formula for the exchange-correlation part. As required by RMTA, spherical muffin-tin potential was used, and semi-relativistic calculation results are presented here. The validity of spherical potential and semi-relativistic approximations were verified by comparing the density of states curve to additionally calculated DOS obtained from the full potential full relativistic KKR method [36], and no significant differences were found (e.g. the total DOS at the Fermi level was 34 $Ry^{-1}$/f.u. versus 31 $Ry^{-1}$/f.u. from semirelativistic muffin-tin calculations). The maximal angular momentum $l_{max} = 4$ was set for all the constituent atoms, calculations were done on a dense k-point mesh (up to 1800 points in the irreducible part of Brillouin zone). The electronic structure was calculated using the experimental crystal structure and lattice parameters, for several external pressure values from 0 to 7.5 GPa (i.e. the range where $T_c$ was measured).

Phonon calculations were done using the plane wave pseudopotential method, as implemented in the Quantum Espresso package [37]. Projector Augmented-Wave (PAW) pseudopotentials were used, with the Generalized Gradient Approximation (GGA) exchange-correlation functional of Perdew, Burke and Ernzerhof [38]. Wave function and charge density cut-offs were set to 48 Ry and 480 Ry, respectively and a k-point mesh of 18x18x18 points in the Brillouin zone was used. The lattice constant relaxation calculations for each pressure were done as a first step, whereby the value for the zero pressure was found to be 6.418 Å, in good agreement with the experimental value of 6.367 Å (see below). For the optimized unit cells, the inter-atomic force constants were obtained by Fourier transformation of the dynamical matrices calculated on a 4x4x4 q-point grid. The phonon DOS was calculated using the tetrahedron integration method and phonon frequencies recalculated to the 10x10x10 q-point mesh. The partial (atomic) phonon DOS were obtained using the QHA package [39].

The combined, electronic structure and phonon calculation results were used to calculate the electron-phonon coupling constant $\lambda_{ep}$ and the superconducting critical temperature, as a function of external pressure.

## III RESULTS AND DISCUSSION

### IIIA. EXPERIMENT

The HfPd$_2$Al as-cast sample was first characterized at room temperature using the powder X-ray diffraction (PXRD) technique. The PXRD confirmed a good quality sample with a very small amount (less than 5%) of HfPdAl impurity phase. The FULLPROF package (ref. [23]) was used to refine a cubic lattice parameter which at room temperature was determined as a = 6.3670(4) Å, close to the value reported in [2,40]. PXRD patterns were then collected above room temperature up to 450$^o$C. The inset of Fig. 1 shows a clear shift of the (422) Bragg peak towards lower angles as the temperature is increased, which is reflected in an increase of the lattice parameter. The thermal expansion in the temperature range 293 K - 723 K is shown in the main panel of Fig. 1. The linear thermal expansion coefficient was found to be $\alpha=1.40(3)\cdot 10^{-5}$ K$^{-1}$, being independent of the temperature in the investigated temperature range. The value is comparable with the result obtained for another Heusler alloy, Ni$_2$MnGa ($\alpha=1.5\cdot 10^{-5}$ K$^{-1}$) [41].

In addition to the PXRD studies, the high pressure behavior of HfPd$_2$Al has been investigated up to 23 GPa by x-ray diffraction as described in Sec. II. The compression data V/V$_0$ as a function of pressure is shown in Figure 2. In order to determine the bulk modulus B$_0$ and its pressure derivative (B$_0$') the data was fitted to the 3$^{rd}$ order Birch-Murnaghan EOS (BMEOS) [42],

$$P = 3B_0 f(1+2f)^{5/2} \times \left[1 + \frac{3}{2}(B_0' - 4)f\right], \quad (2)$$

With $f = \frac{1}{2}\left[(V_0/V)^{2/3} - 1\right]$, which yielded B$_0$ = 97(2) GPa and B'$_0$ = 8.5(5). The extrapolated cell parameter value at ambient conditions (room pressure) was found to be 6.3712(35) Å.

Figure 3a presents the temperature dependence of the heat capacity divided by temperature (C$_p$/T) near to the superconducting transition. A sharp anomaly at T$_c$ = 3.53 K confirms the bulk nature of the superconductivity and the good quality of the tested HfPd$_2$Al sample. The T$^2$ dependence of C$_p$/T, measured at a magnetic field of μ$_0$H = 3 T, which exceeds H$_{c2}$, is shown in Fig. 3b. A curve through the data points shows the fit of C$_p$/T = γ + βT$^2$ + δT$^4$ in the temperature range 2 K < T < 5 K. The fit reveals a Sommerfeld coefficient γ = 7.6(3) mJ mol$^{-1}$ K$^{-2}$ and Debye temperature Θ$_D$ = 177(3) K (where $\beta = \frac{12\pi^4 N k_B}{5\Theta_D^3}$ and N = 4 is the number of atoms per formula unit). Knowing the γ value, one can calculate the specific heat jump ΔC/γT$_c$ = 1.59 which is very close to the value reported in ref [2] ΔC/γT$_c$ = 1.5. A logarithmic averaged phonon frequency can be determined from $\frac{\Delta C}{\gamma T_c} = 1.43\left[1 + 53\left(\frac{T_c}{\omega_{\log}}\right)^2 \ln\left(\frac{\omega_{\log}}{3T_c}\right)\right]$ and for HfPd$_2$Al we obtained ω$_{\log}$ = 120 K. The electron phonon coupling constant λ$_{ep}$ can also be estimated from the inverted Allen and Dynes equation for T$_c$ [43] (see, Eq. 9):

$$\lambda_{ep} = \frac{1.04 + \mu^* \ln(\omega_{\log}/1.2T_c)}{(1 - 0.62\mu^*)\ln(\omega_{\log}/1.2T_c) - 1.04} \quad (3)$$

Taking T$_c$ = 3.53 K, ω$_{\log}$ = 120 K and a Coulomb repulsion constant μ$^*$ = 0.1, we obtained λ$_{ep}$ = 0.657 which confirms that HfPd$_2$Al is a moderately coupled superconductor. Note, that the Allen-Dynes pre-factor ω$_{\log}$/1.2 in T$_c$ Eq. (9) was originally fitted [43] using a lower value of the Coulomb pseudopotential parameter, μ$^*$ = 0.1, than the "standard" McMillan's μ$^*$ = 0.13 (see, also Ref. [44]), so we consequently used $\mu^* = 0.1$ whenever Eq. (3) or Eq. (9) were used, otherwise Eq. (9) underestimates the $T_c$ comparing to McMillans equation.

The superconducting transition was further examined through temperature dependent measurements of the electrical resistivity under applied pressure from p = 0 GPa to 7.49 GPa. Normalized resistivity $\rho(T) / \rho(4K)$ is shown in panels (a) and (b) of Figure 4. As can be seen from Fig. 4a, under ambient pressure the superconducting transition is sharp and $T_c$ = 3.55 K. This value is in very good agreement with the $T_c$ determined from heat capacity measurements, and slightly lower than reported in ref. [2]. A slightly higher $T_c$ and a much broader transition is observed for $HfPd_2Al$ under applied pressure (see Fig. 4b). The first effect is likely caused by decreased electrical current (I=0.5 mA) used during measurements reducing Ohm effect in the sample while the second one is due to quasi hydrostatic pressure conditions of the solid tramsmitting medium. The superconducting critical temperature, $T_c$, was defined as the temperature at which $R(T)/R(4K) = 0.5$. These data correspond to the open circles in the panel (c) of Figure 4 that presents the pressure dependence of $T_c(p)$. The superconducting critical temperature monotonically decreases with the pressure increase and the slope $dT_c/dp$ = -0.13(1) K GPa$^{-1}$. Neglecting the potential influence of applied pressure on a logarithmic averaged phonon frequency (discussed in the theoretical part), we took $\omega_{log}$ = 120 K and calculated the electron-phonon coupling constant ($\lambda_{ep}$) for each value of applied pressure by using Eq. 3. As can be seen from Fig. 4d, $\lambda_{ep}(p)$ decreases linearly from approximately 0.67 to 0.6, estimated for p = 0.05 GPa and 7.49 GPa, respectively.

The electron-phonon coupling may also be estimated from the inverted McMillan formula [29]:

$$\lambda_{ep} = \frac{1.04 + \mu^* \ln\left(\frac{\theta_D}{1.45T_C}\right)}{(1 - 0.62\mu^*)\ln\left(\frac{\theta_D}{1.45T_C}\right) - 1.04} \quad (4)$$

For ambient pressure, $T_c$ = 3.55 K and $\Theta_D$ = 177 K, we obtained $\lambda_{ep}$ = 0.68 and 0.61 for $\mu^*$ = 0.13 and 0.1, respectively [45]. These values are very close to $\lambda_{ep}$ = 0.65 based on the heat capacity measurement and Allen-Dynes formula.

**IIIB. THEORY**

The electron-phonon coupling parameter, $\lambda_{ep}$, can be approximately calculated using the formula [31]:

$$\lambda_{ep} = \sum_i \frac{\eta_i}{M_i \langle \omega_i^2 \rangle} \quad (5)$$

where $\eta_i$ is the McMillan-Hopfield parameter of the $i$-th atom in the unit cell, with mass $M_i$ and average square phonon frequency $\langle \omega_i^2 \rangle = \int \omega F(\omega) d\omega / \int \omega^{-1} F(\omega) d\omega$, and $F(\omega)$ is the phonon DOS. In this way $\lambda_{ep}$ is a sum of the contributions from each atoms' sublattice (in our case these are Hf, Pd with 2 atoms, and Al).

The pressure dependence of $T_c$ is mainly determined by the pressure dependence of $\lambda_{ep}$ (the increase in $\omega_{log}$ or $\Theta_D$ under pressure is less important). In multi-atomic crystals, for each of the sublattices (or for the monoatomic system), this is controlled by the ratio of electronic and phonon contributions, $\eta_i$ and $\langle \omega_i^2 \rangle$. Since $\langle \omega_i^2 \rangle$ describes stiffness of the lattice and is expected to increase with pressure, the change of the McMillan-Hopfield parameter $\eta_i$ is the key factor determining the response of superconductor to external pressure. For each sublattice we may calculate the logarithmic derivative:

$$\frac{d \ln \lambda_i}{dP} = -\frac{1}{\tilde{B}} \left( \frac{d \ln \eta_i}{d \ln V} - \frac{d \ln \langle \omega_i^2 \rangle}{d \ln V} \right), \quad (6)$$

where $P$ is pressure, $V$ is the unit cell volume, and $\tilde{B}$ is the bulk modulus defined by the simplified volume-pressure dependence equation $V(P) = V_0 \exp(-P/\tilde{B})$, used to convert the pressure derivative into the volume one. To avoid confusions with the bulk modulus determined from the Birch-Murnaghan EOS (Eq. 2), the symbol $\tilde{B}$ is used here. In the pressure range 0-7.5 GPa the fit of the experimental $V(P)$ data to this equation gives $\tilde{B} \approx 124$ GPa, however its value is not important for the sign of the pressure dependence of $\lambda_{ep}$, which we analyze here. Now, defining the average Grüneisen parameter as:

$$\gamma_G = -\frac{d \ln \langle \omega \rangle}{d \ln V} \approx -\frac{1}{2} \frac{d \ln \langle \omega^2 \rangle}{d \ln V} \quad (7)$$

the simple rule can be obtained for the pressure effect on EPC. $\lambda_i$ decreases with pressure as long as the term in brackets in Eq. (6) is positive, i.e. when $\frac{d\ln\eta_i}{d\ln V} > -2\gamma_G^i$.

Since $\frac{d\ln\eta_i}{d\ln V}$ is usually negative [31, 46] and $\gamma_G$ positive, it is convenient to write it as:

$$-\frac{d\ln\eta_i}{d\ln V} < 2\gamma_G^i, \qquad (8)$$

i.e. behavior of $\lambda_i$ with pressure is determined by the magnitude of the Grüneisen parameter and logarithmic derivative of the McMillan-Hopfield parameter. $\lambda_i$ decreases if the increase of $\eta_i$ with decreasing unit cell volume is slower than the lattice stiffening, described by the double $\gamma_G$ parameter of the sublattice $i$.

The electronic structure and phonon calculations were undertaken to verify whether $T_c$ and its pressure dependence in $HfPd_2Al$ can be described within this conventional scenario. Figure 5, top panel, shows the electronic density of states of $HfPd_2Al$ at ambient pressure, with partial atomic densities marked by colors. Bottom panel of Fig. 5 compares DOS near the Fermi level for 0 and 7.45 GPa external pressures. The highest contribution to the DOS near $E_F$ comes from the two Pd atoms' 4d states. The corresponding electronic bands are plotted for the same set of pressures in Fig. 6. As was pointed out in ref. [40], the Fermi level in this family of Heusler compounds lies between van Hove singularities, with the closest singularity located at the L point, which is shown well in Fig. 6. Upon applying external pressure, the location of this singularity remains almost unchanged. Table I presents the computed electronic structure parameters of $HfPd_2Al$. The Hf and Pd atoms contribute equally to the density of states at $E_F$, if counting per atom, with a minor contribution from Al. As pressure increases, due to the increased hybridization, values of n($E_F$) slightly decrease. The McMillan-Hopfield $\eta_i$ parameter is highest for the Pd atom, with a dominating contribution from the *d-f* scattering channel, as typical for the transition metal element [31, 47]. A slightly smaller value of $\eta_i$ is found on Hf, and here *p-d* and *d-f* contributions are equally important. When the unit cell volume decreases, $\eta_i$ for Hf and Pd are increasing, as a result of the increase in the matrix element of the potential gradient between the radial wave functions of *l* and *l+1* type (see, Eq. 1). The opposite tendency is found for Al - here $\eta_i$ is gradually decreasing with pressure. The modifications of $\eta_i$ in the pressure range 0-7.45 GPa are plotted in Fig. 7, and the slopes of the logarithmic derivatives versus volume of the primitive cell (V), $\frac{d\ln\eta_i}{d\ln V}$, calculated by linear fitting of the $\ln\eta_i$ versus $\ln V$, are reported in Table II, and are described later.

The phonon densities of states and phonon dispersions are shown in Fig. 8 and Fig. 9, respectively, with the partial atomic DOS plotted in colors in Fig. 8. Due to large differences in atomic masses ($M_{Hf} \approx 178$ u, $M_{Pd} \approx 106$ u, $M_{Al} \approx 27$ u), a gap in the phonon spectrum is formed, with the high frequency range due to aluminum oscillations. Although hafnium is heavier than palladium, it seems from our calculations that the density of states around the frequency of 4 THz have larger populations for this atom, resulting in smaller $\langle \omega_i^2 \rangle$. When pressure is increased to 7.45 GPa, the vibration spectrum moves towards higher frequencies as also observed in the phonon dispersions plot. Generally, the phonon dispersions are only shifted towards higher frequencies with pressure, except for the lowest acoustic mode in the K-Γ direction. We observe the softening of this mode, with the minimum value decreasing from 0.7 THz at P = 0 GPa to 0.626 THz for P = 7.45 GPa. This suggests the possibility of a structural transition of HfPd$_2$Al at higher pressures. Also, similar soft-mode behavior, but already evident at ambient pressure, as indicated by the imaginary frequency around the minimum point, was previously reported in ref. 40 from calculations in the isoelectronic compound ZrPd$_2$Al. Thus the observed soft mode behavior may be a more general property of this group of Heusler alloys. Phonon anomalies were also recently observed in superconducting YPd2Sn Heusler compound[48].

Now returning to the analysis of electron-phonon coupling, the average square phonon frequencies, calculated using partial phonon densities for 0 and 7.45 GPa are collected in Table I. Using these values and corresponding McMillan-Hopfield parameters, the atomic contributions, $\lambda_i$, to the electron-phonon coupling constant, $\lambda_{ep}$, are calculated and presented in that Table. The $\lambda_i$ coefficient calculated per atom in the unit cell is greatest for palladium, being more than twice the value for hafnium. Together with the fact, that there are two Pd atoms in the primitive cell of HfPd$_2$Al, the electron-phonon coupling value is mainly the effect of interaction between electrons and phonons on palladium atoms. The aluminum contribution to the total EPC is found to be negligible. The calculated values of total $\lambda_{ep}$, average phonon frequency and critical temperature $T_c$ as function of pressure for four pressure values (0, 2.53, 4.50, and 7.45 GPa) are collected in Table III. Critical temperature is calculated from the Allen-Dynes formula [43]:

$$T_c = \frac{\omega_{\log}}{1.20} \exp\left[\frac{-1.04(1+\lambda_{ep})}{\lambda - \mu^*(1+0.62\lambda_{ep})}\right] \quad (9)$$

where:

$$\omega_{\log} = \exp\left[\frac{\int F(\omega) \ln \omega \frac{d\omega}{\omega}}{\int F(\omega) \frac{d\omega}{\omega}}\right] \quad (10)$$

Note, that the theoretical value of $\omega_{\log}$ = 110 K (Table III) at ambient pressure corresponds very well with the value estimated from the heat capacity measurement $\omega_{\log}$ = 120 K, discussed earlier.

From Table III we see, that the zero-pressure value of $\lambda_{ep}^{calc}$ = 0.54 is about 20% lower, than the "experimental" value deduced from $T_c$ for "as-cast" samples ($\lambda_{ep}^{exp}$ = 0.66). Possible reasons for this underestimation may be both 'technical' (like inaccuracy of the RMTA, which is known to underestimate EPC in some materials, see, e.g. Ref. [32], or neglecting of the off-diagonal terms in Eq. (5), i.e. setting $\lambda_{ij} = \lambda_i \delta_{ij}$) and 'chemical', i.e. assumption that the investigated sample is a defect-free and perfectly ordered crystal. The latter assumption is likely not to be true, due to the observed dependence of $T_c$ on the heat treatment applied to the sample, i.e. decreasing of $T_c$ after annealing [2]. We actually expect a significant amount of anti-site defects to be present there (anti-site disorder was observed in many related Heusler and half-Heusler systems, like $Co_2MnGe$ [49], $Fe_{1-x}Ni_xTiSb$ [50] and FeVSb [51]). Moreover, our preliminary KKR calculations with the coherent potential approximation (CPA) for the system with anti-site defects showed, that EPC is likely to increase if anti-site defects are present. This could potentially explain the underestimation of $\lambda_{ep}$ but this subject is beyond the scope of this paper and will be addressed in future studies. The 20% underestimation in $\lambda_{ep}$, due to the exponential dependence of $T_c$ on $\lambda_{ep}$, gives a much larger underestimation of $T_c$. Nevertheless, the decreasing tendency for both $T_c$ and $\lambda_{ep}$ is well reproduced in the presented calculations, as shown in Fig. 10, where the relative changes are plotted. The analysis of the data in Table II explains the reason for the decrease in $\lambda_{ep}$, in agreement with the description of Eq. 8. For both, Pd and Hf sublattices, the condition $-\frac{d\ln\eta_i}{d\ln V} < 2\gamma_G^i$ is fulfilled, i.e. the lattice stiffening occurs faster than the increase in the McMillan-Hopfield parameters, resulting in a decreasing of the EPC strength. The calculated value of the "average" Grüneisen parameter, obtained from the total phonon spectrum in the 0 - 7.45 GPa pressure range, is $\gamma_G$ = 1.87 ± 0.22. This is a quite typical value, and similar ones were reported for other Heusler alloys such as $Ni_2MnSn$ ($\gamma_G$ = 1.86 [52]), $Ti_2FeGe$ ($\gamma_G$ = 2.44) and $Ti_2FeSn$ ($\gamma_G$ = 2.50) [53]. This is not the case for other examples such as $Ni_2MnGe$ or $Ni_2MnSb$ which have much lower values, $\gamma_G$ = 0.21 and 0.47, respectively [52]. To verify the correctness of our calculated $\gamma_G$, we have estimated the Grüneisen parameter using the experimental results, based on the formula [54]:

$$\gamma_G = \frac{3\alpha K_T}{C_V \rho}.$$

Here, $\alpha$ is the linear thermal expansion coefficient, $K_T = -V\frac{\partial P}{\partial V}\bigg|_T$ is the isothermal bulk modulus, $C_V$ is the constant volume heat capacity (per mass) and $\rho$ is the mass density.

The pressure-dependent $K_T$ values at 300 K were directly computed from the fitted BMEOS (see, above), and the constant volume heat capacity was taken from the Dulong-Petit law (calculations of the specific heat, using the theoretical phonon DOS, showed, that for $HfPd_2Al$ at 300 K $C_v$ is already 96% of the Dulong-Petit value of 12R/mole/f.u., so the difference is negligible). The resulting $\gamma_G$ was equal 1.6 at low pressures, so in reasonable agreement with the theoretical value $\gamma_G$ = 1.87. Values for higher pressures, calculated by neglecting the pressure dependence of $\alpha$ (so less accurate) are increasing via $\gamma_G$ = 2.0 at 3 GPa to $\gamma_G$ = 2.5 at 7.45 GPa, due to an increase of $K_T$. We observe a smaller theoretical Grüneisen parameter at higher pressures, as well as a larger theoretical value of $\tilde{B}_{theor} \approx 173$ GPa, fitted using equation $V(P) = V_0 \exp(-P/\tilde{B})$. This helps to explain the slower decrease of $\lambda_{ep}$ and $T_c$ with pressure at higher pressures, compared to that observed experimentally, as plotted in Fig. 10.

## IV. SUMMARY AND CONCLUSIONS

In summary, $HfPd_2Al$ a polycrystalline sample was synthesized by the arc-melting method and its purity checked by powder x-ray diffraction. The heat capacity and electrical resistivity measurements confirm bulk superconductivity. Estimated superconducting critical temperature, Sommerfeld parameter, Debye temperature and the heat capacity superconducting jump are in good agreement with those previously reported for $HfPd_2Al$ [2].

Our high temperature xrd study of $HfPd_2Al$ reveals the linear thermal expansion coefficient $\alpha = 1.40(3) \cdot 10^{-5}$ $K^{-1}$, which is comparable with the result obtained for $Ni_2MnGa$ ($\alpha=1.5 \cdot 10^{-5}$ $K^{-1}$) [41]. The main part of this paper is focused on the compression and resistivity data under applied pressure. The estimated bulk modulus for $HfPd_2Al$ at room temperature is $B_0 = 97(2)$ GPa, which is much smaller than the value obtained from the electronic structure calculations (B=159 GPa) [40]. The superconducting critical temperature decreases linearly with applied pressure and $dT_c/dp = -0.13(1)$ K $GPa^{-1}$. The negative slope of $T_c(p)$ is contrary to expectations based on the lattice parameter $T_c(a)$ dependence observed for (Hf, Zr)$Pd_2$(In, Al) Heusler alloys, for which the critical temperature increases with decreasing lattice parameter *a*. Thus, the differences in $T_c$ among (Hf, Zr)$Pd_2$(In, Al) Heuslers likely come from the differences in electronic and phonon structures and are not just the effect of chemical pressure. The value of $dT_c/dp$ estimated for $HfPd_2Al$ is comparable to the value obtained for $ScPd_2Al$ ($dT_c/dp = -0.145$ K $GPa^{-1}$) and remains the lowest among those reported for $APd_2Sn$ (A=Sc, Y, Tm, Yb, Lu) and $YPd_2Pb$ Heusler type superconductors [15].

Electronic structure and phonon calculations were performed on $HfPd_2Al$ for several external pressures in the range 0 - 7.5 GPa. Using the Rigid Muffin Tin Approximation an electron-phonon coupling constant $\lambda_{ep}$ was calculated. We found that the highest contribution to $\lambda_{ep}$ in $HfPd_2Al$ comes from the Pd atoms sublattice. Under external pressure the electronic part of the EPC constant, i.e. McMillan-Hopfield parameters $\eta$, increase for Pd and Hf and decrease for Al. Nevertheless, the pressure induced stiffening of the crystal lattice, represented by the average Grüneisen parameter $\gamma_G = 1.87$ (calculations), $\gamma_G = 1.60$ (experiment), overcomes the increase in $\eta_{Pd}$ and $\eta_{Hf}$, thus the total electron-phonon coupling constant $\lambda_{ep}$ decreases. Although the initial (zero pressure) value of the electron-phonon coupling parameter $\lambda_{ep} = 0.54$ is underestimated in calculations by about 20%, compared to the experimental value $\lambda_{ep}$ = 0.67, the theoretical results explain rather well the experimentally observed decrease in $\lambda_{ep}$ and $T_c$ with pressure which are a result of considerable lattice stiffening and not compensated enough by an increase of the McMillan-Hopfield parameters.

**Acknowledgments**

BW was supported by the National Science Center (Poland) (project no. DEC-2011/02/A/ST3/00124). The research performed at Gdansk University of Technology was financially supported by the National Science Centre (Poland) grant (DEC-2012/07/E/ST3/00584).

Table I. Site-decomposed electronic and dynamic properties of HfPd$_2$Al for $P = 0$ and 7.45 GPa. $n_i(E_F)$ is in Ry$^{-1}$/spin, $\eta_i$ in mRy/a$_0^2$ (both per atom, a$_0$ is the atomic Bohr radius), $\omega_i$ in THz.

| atom | $n_i(E_F)$ | $n_s(E_F)$ | $n_p(E_F)$ | $n_d(E_F)$ | $n_f(E_F)$ | $\eta_i$ | $\eta_{sp}$ | $\eta_{pd}$ | $\eta_{df}$ | $\sqrt{\langle\omega_i^2\rangle}$ | $\lambda_i$ |
|---|---|---|---|---|---|---|---|---|---|---|---|
| colspan="12" | $P = 0$ GPa |
| Hf | 4.0 | 0.03 | 0.17 | 3.74 | 0.021 | 10.74 | 0.05 | 4.94 | 5.75 | 2.85 | 0.088 |
| Pd | 4.0 | 0.40 | 0.70 | 2.88 | 0.038 | 11.7 | 0.86 | 2.83 | 8.01 | 2.46 | 0.223 |
| Al | 1.7 | 0.14 | 1.35 | 1.66 | 0.010 | 0.67 | 0.03 | 0.62 | 0.02 | 5.59 | 0.009 |
| colspan="12" | $P = 7.45$ GPa |
| Hf | 3.5 | 0.03 | 0.16 | 3.28 | 0.020 | 12.24 | 0.08 | 6.23 | 5.93 | 3.08 | 0.086 |
| Pd | 3.7 | 0.37 | 0.60 | 2.71 | 0.040 | 14.41 | 1.34 | 3.09 | 9.97 | 2.77 | 0.210 |
| Al | 1.5 | 0.12 | 1.21 | 1.50 | 0.010 | 0.58 | 0.10 | 0.45 | 0.02 | 6.14 | 0.007 |

Table II. First row: slope of the logarithmic derivative of the McMillan-Hopfield parameters versus volume of the primitive cell (V), calculated by linear fitting of $\ln \eta_i$ vs $\ln V$. Second row: atomic Grüneisen parameter, calculated as a slope of $\ln \langle \omega_i \rangle$ vs $\ln V$, where $\langle \omega_i \rangle$ is the partial average phonon frequency. Third row: same as second, but for the total phonon spectrum. Values in brackets are standard deviations of fittings.

|  | Hf | Pd | Al |
|---|---|---|---|
| $\dfrac{d \ln \eta_i}{d \ln V}$ | -2.16(0.13) | -3.28(0.22) | 2.08(0.32) |
| $\gamma_G^i$ | 1.40(0.07) | 2.02(0.09) | 1.53(0.06) |
| $\gamma_G$ (total) | 1.87(0.22) |||

Table III. Theoretical results for $\lambda_{ep}$, $\omega_{log}$, $<\omega>$ and $T_c$ for investigated pressures. $\omega_{log}$ and $<\omega>$ increase with pressure in about 10%, however drop in $\lambda_{ep}$ is strong enough to compensate this effect and decrease in $T_c$ with pressure is obtained from calculations.

|  | 0 GPa | 2.25 GPa | 4.53 GPa | 7.45 GPa |
|---|---|---|---|---|
| $\lambda_{ep}$ | 0.543 | 0.528 | 0.520 | 0.512 |
| $\omega_{log}$ | 110 K | 116 K | 118 K | 121 K |
| $<\omega>$ | 187 K | 195 K | 200 K | 208 K |
| $T_c$ | 1.81 K | 1.73 K | 1.67 K | 1.61 K |

**Figures**

Figure 1. (color online)

Relative change of the *a* lattice parameter of HfPd$_2$Al with increasing temperature. The data were estimated by the LeBail method using the FULLPROF package. Inset shows a clear shift of the (422) Bragg peak of HfPd$_2$Al towards lower angles as temperature is increased.

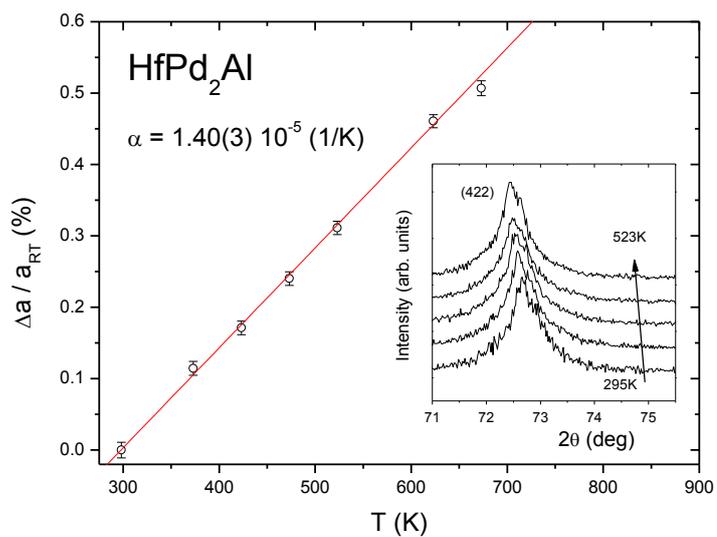

Figure 2. (color online)

Variation of relative volume with pressure for HfPd$_2$Al. The blue curve is the fit to the 3$^{rd}$ order Birch-Murnaghan EOS (BMEOS). For more details see text.

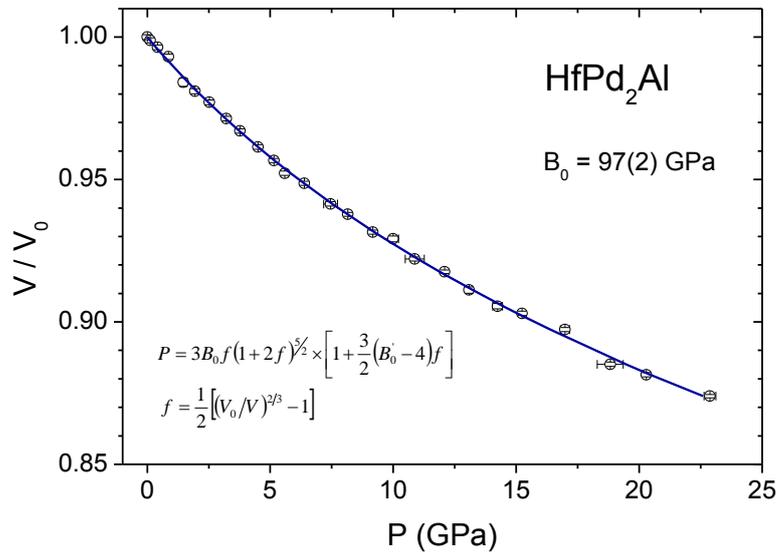

Figure 3. (color online)

Panel (a): temperature dependence of heat capacity divided by temperature ($C_p/T$) of HfPd$_2$Al measured in zero magnetic field in the vicinity of the superconducting transition. Panel (b): $C_p/T$ versus $T^2$ measured in $\mu_0 H = 3T$. The red curve is a fit of $C_p/T = \gamma + \beta T^2 + \delta T^4$.

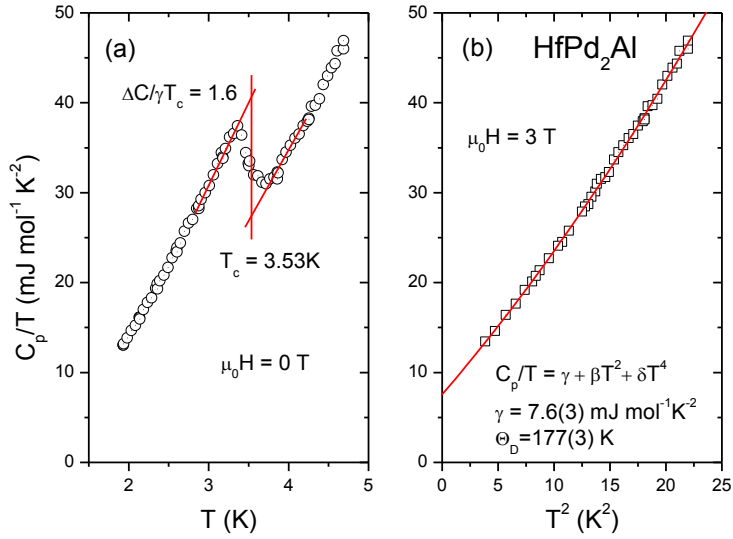

Figure 4. (color online)

Normalized electrical resistivity under ambient pressure (a) and various applied pressures (b) of HfPd$_2$Al. Pressure dependence of the superconducting critical temperature (c) and electron-phonon coupling parameter (d).

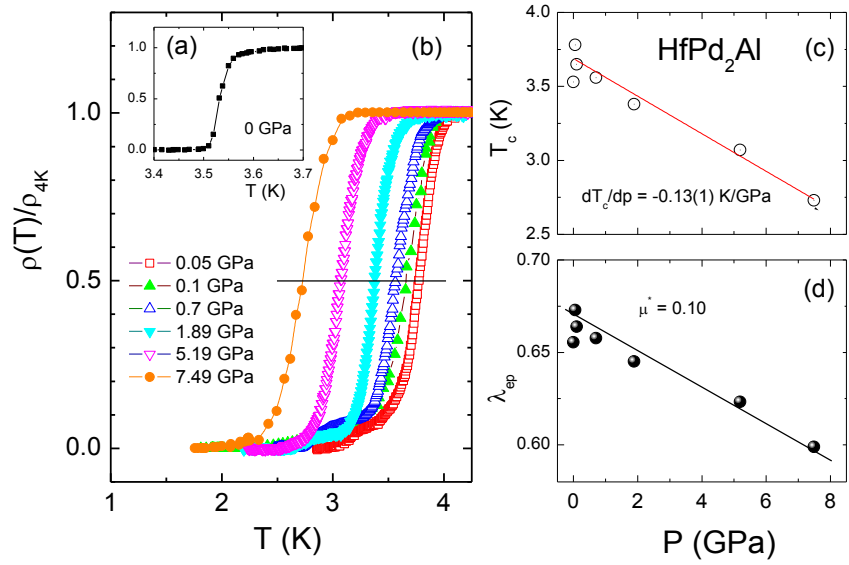

Figure 5. (color online)

Top panel: Electronic density of states for HfPd$_2$Al at ambient pressure. Bottom panel: comparison of DOS near the Fermi level for 0 and 7.45 GPa external pressures. Color lines show the atomic partial DOS (per atom).

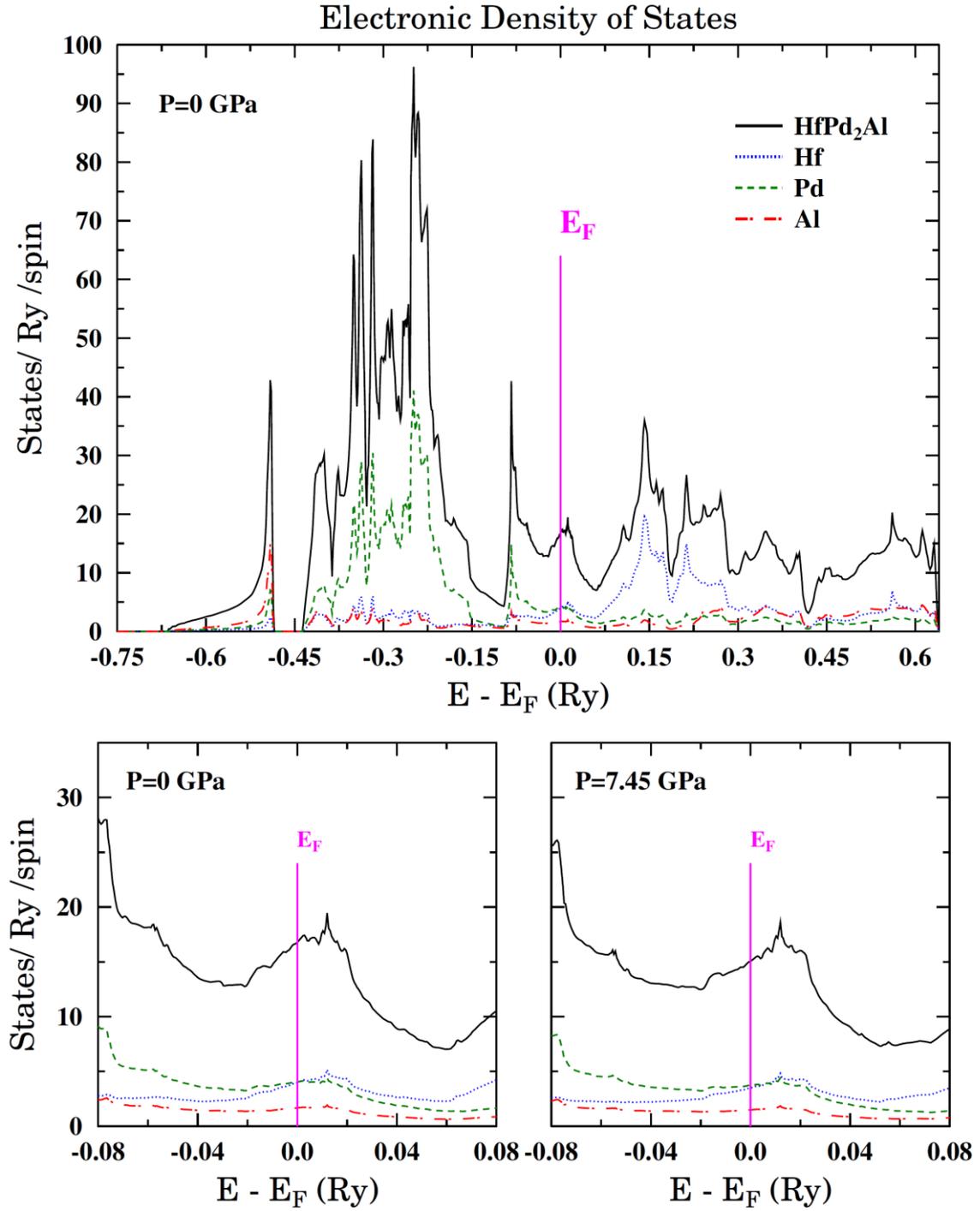

Figure 6. (color online)

Top panel: Electronic dispersion relations for HfPd$_2$Al at ambient pressure. Bottom panel: comparison of bands near the Fermi level for 0 and 7.45 GPa external pressures.

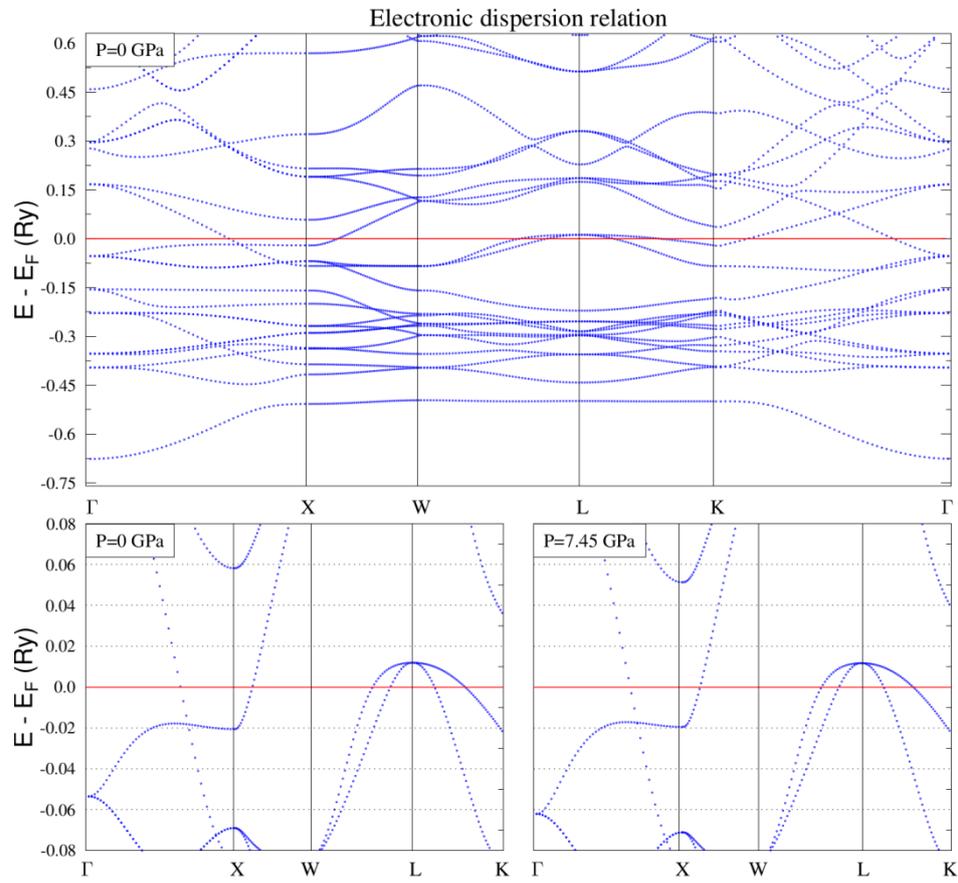

Figure 7. (color online)

Effect of pressure on McMillan-Hopfield parameters.

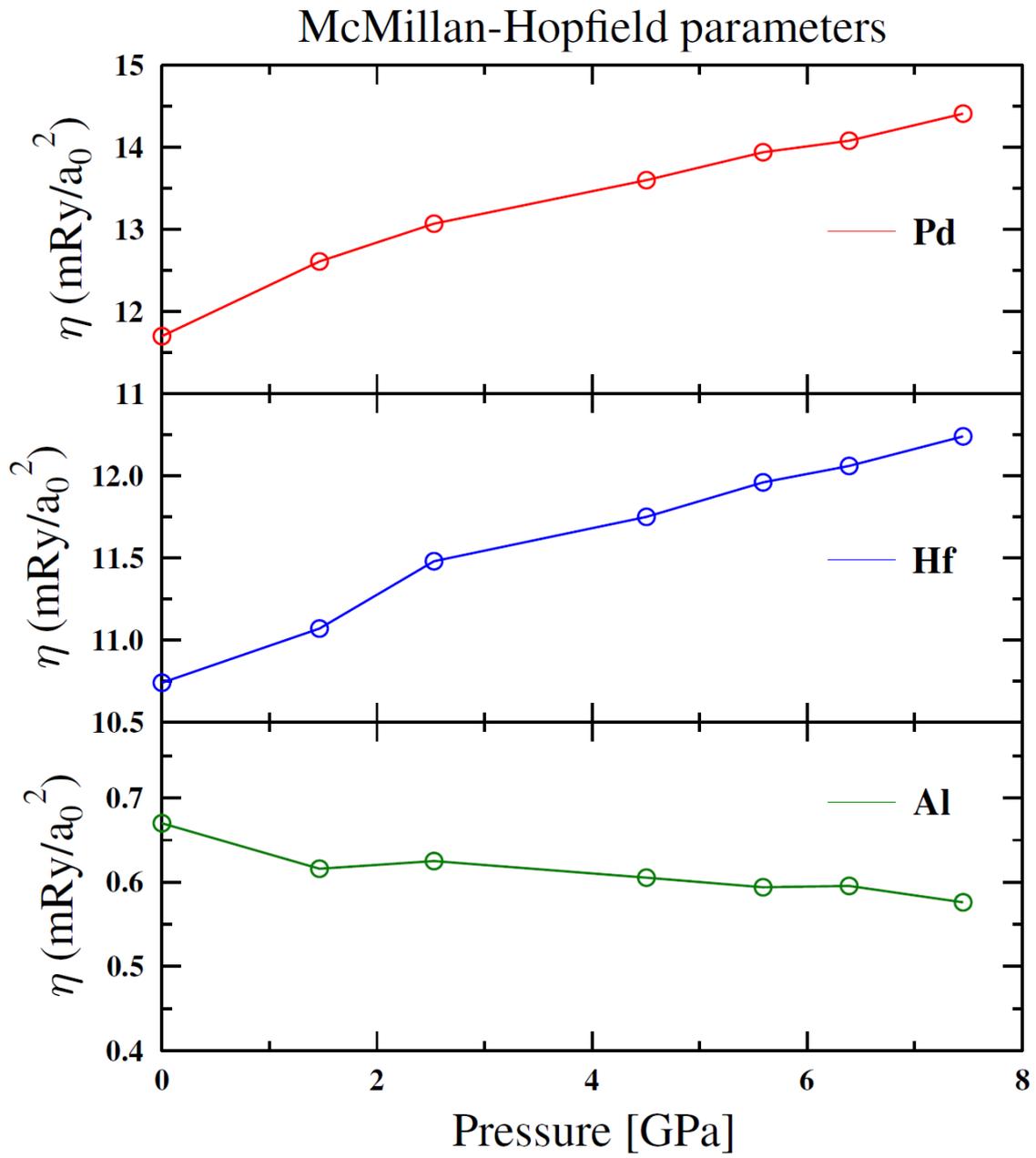

Figure 8. (color online)

Phonon density of states for HfPd$_2$Al at 0 (top panel) and 7.45 GPa (bottom panel). Color lines show the atomic partial DOS (per atom).

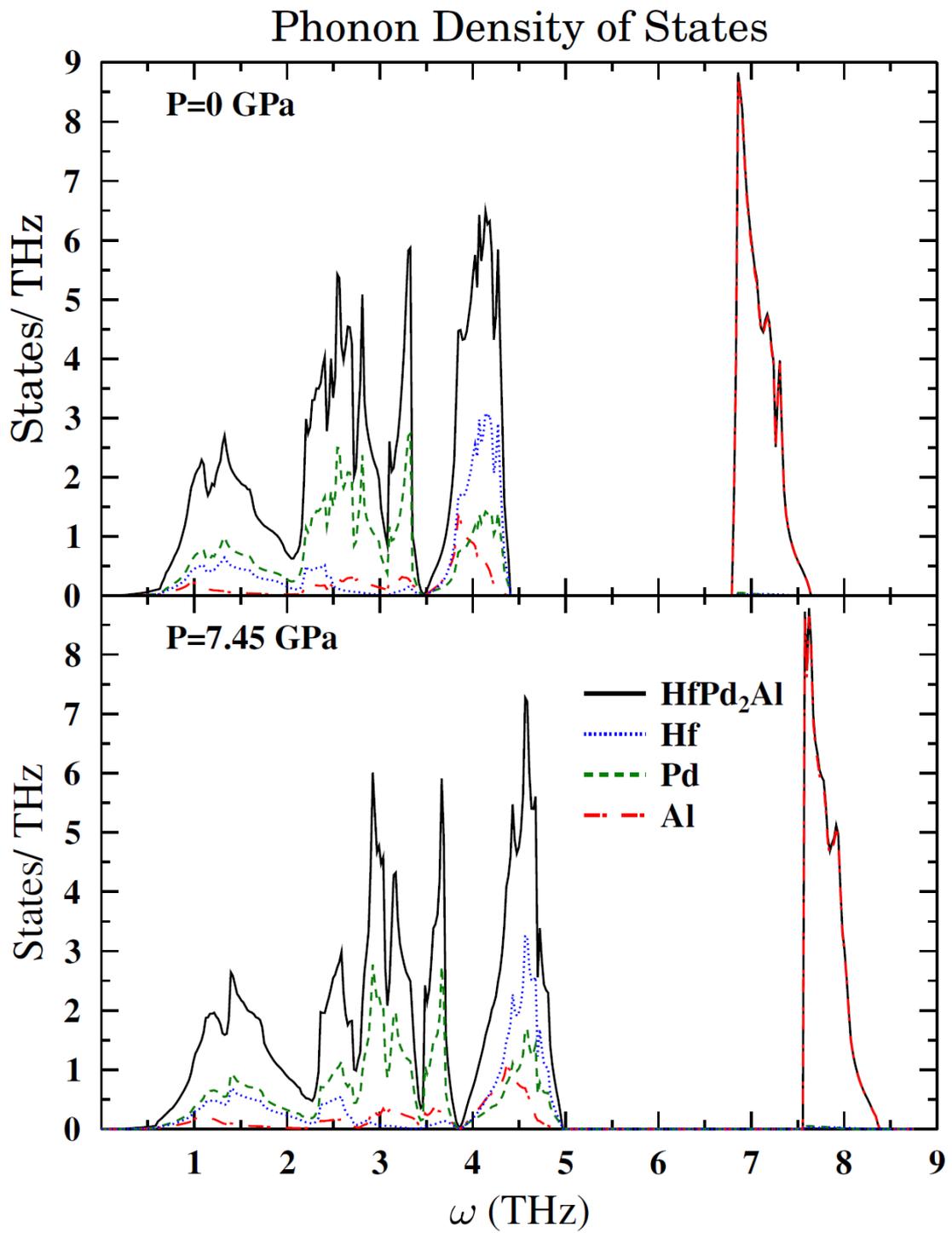

Figure 9 (color online)

Phonon dispersion relations for HfPd$_2$Al at 0 (top panel) and 7.45 GPa (bottom panel).

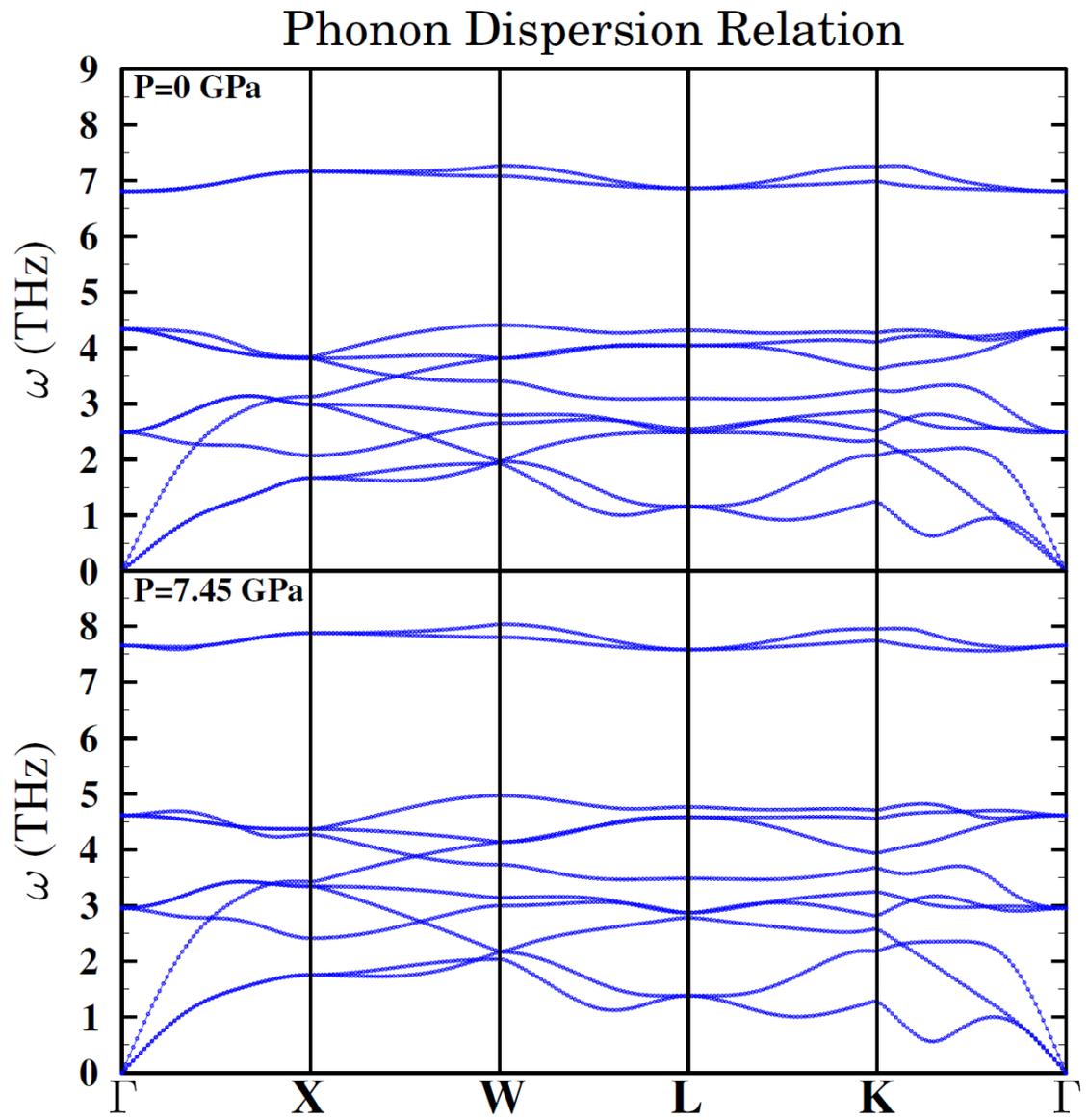

Figure 10. (color online)

Relative changes of electron-phonon coupling parameter $\lambda_{ep}$ (top panel) and critical temperature $T_c$ (bottom panel) obtained from calculations (empty symbols) and measurements (full symbols). The overall trends are very similar, but theoretical results underestimate the experimentally observed changes for the highest pressures.

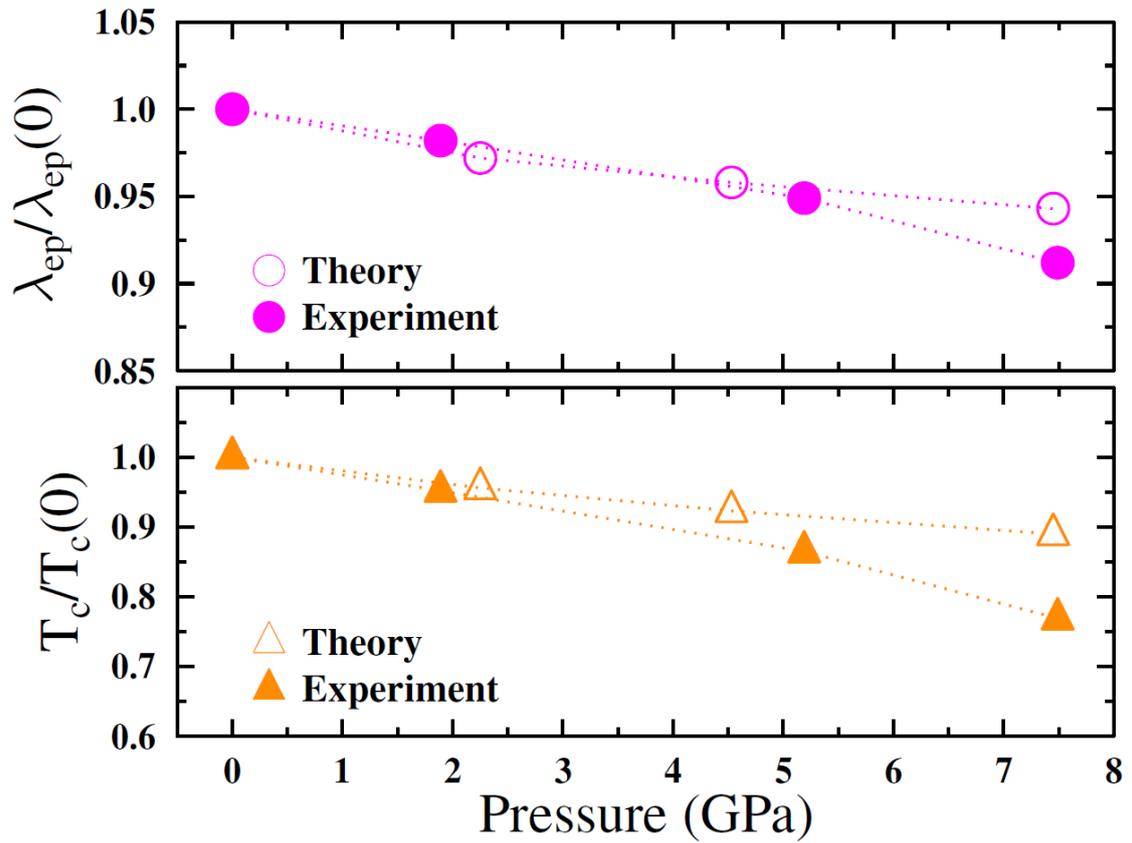